\documentclass[preprint,12pt, nofootinbib]{revtex4}

\usepackage[brazil, english]{babel}
\usepackage[utf8]{inputenc}
\usepackage{graphicx}
\usepackage{epsfig}
\usepackage{amssymb}
\usepackage{amsmath} 
\usepackage{slashed}
\usepackage[unicode=true,bookmarks=false,breaklinks=false,pdfborder={0 0 1},colorlinks=true]
 {hyperref}
\hypersetup{
 citecolor=blue,linkcolor=blue,urlcolor=blue}

\graphicspath{%
    {converted_graphics/}
    {/}
}

\usepackage{tikz,tikz-3dplot}
\tdplotsetmaincoords{80}{45}

\tikzset{surface/.style={draw=black, fill=white, fill opacity=.6}}

\usepackage{tikz}
\usetikzlibrary{decorations.pathmorphing,patterns}
\begin{document}

\title{From Einstein to Horndeski: Holographic Transport Coefficients in Modified Gravity}

\author{Fabiano F. Santos$^{1}$}
\email[Eletronic address: ]{fabiano.ffs23@gmail.com}
 
\affiliation{$^1$Departamento de Física, Universidade Federal do Maranhão, Campus Universitario do Bacanga, São Luís (MA), 65080-805, Brazil.}

\begin{abstract}
This paper examines the holographic computation of bulk and shear viscosity ratios in strongly coupled thermal plasmas using the AdS/BCFT correspondence within Horndeski gravity. We demonstrate that this framework leads to non-zero viscosity-to-entropy ratios ($\zeta/S$ and $\eta/S$) at low temperatures, indicating a break in conformal symmetry. At high temperatures, these ratios approach zero, recovering the expected conformal behavior of quark-gluon plasma. Our findings provide new insights into the hydrodynamic properties of strongly coupled plasmas and offer a more nuanced understanding of QCD-like theories in holographic models.
\end{abstract}
 

\maketitle

\newpage
\section{Introduction}

In recent years, the macroscopic properties of strongly coupled matter have been widely discussed and still have open problems/questions due to the need for non-perturbative methods \cite{Skokov:2009qp,Fukushima:2016xgg,Meyer:2007dy,Kharzeev:2007wb}. A system that is still the scene of great discussion both from a theoretical and experimental point of view is the quark-gluon plasma (QGP) \cite{Ahn:2022azl,Ballon-Bayona:2022uyy,Gubser:2008px,Gubser:1996de,Burgess:1999vb,Policastro:2001yc,Policastro:2002se,Buchel:2007mf}, which is produced in heavy ion collisions at the Relativistic Heavy Ion Collider (RHIC) and the Large Hadron Collider (LHC); the reason for this is due to the fact that experimentalists \cite{Skokov:2009qp,Fukushima:2016xgg} during this collision find a small bulk viscosity suggesting that this strongly coupled system is closer to "a fluid than a plasma."

Many investigations into holographic transport coefficients, such as shear and bulk viscosities, have been carried out during the last years; these techniques only require the dependence of the leading frequency on an appropriate Green's function in the low-frequency limit. Thus, using Einstein gravity coupled with a scalar field, which mimics the QCD equation of state, a slight violation of the lower bound on the ratio of shear and bulk viscosities was observed both in Einstein gravity models \cite{Kharzeev:2007wb,Buchel:2007mf,Karsch:2007jc,Gubser:2008yx,Gubser:2008sz} and beyond \cite{Bravo-Gaete:2020lzs,Bravo-Gaete:2022lno,Santos:2023flb,Bravo-Gaete:2023iry,Santos:2023mee}. Recent investigations have shown that the bulk viscosity ($\zeta$) can undergo an increase near a critical temperature of Quantum Chromodynamics (QCD); this regime is where confinement occurs and chiral symmetry breaking sets in \cite{Kharzeev:2007wb,Karsch:2007jc}.

Investigations by \cite{Gubser:2008yx,Gubser:2008sz} showed that Einstein gravity coupled with a scalar field increases the bulk viscosity near a critical temperature (T$_c$). However, this increase is not as significant as expected for the pure Yang-Mills theory \cite{Meyer:2007dy,Kharzeev:2007wb}. On the other hand, numerical results presented by \cite{Gubser:2008yx} led to the interpretation that $\zeta/S$ diverges in the holographic scenario in the two-derivative supergravity approximation; the entropy density $S$ has an extremum as a function of temperature.

\begin{figure}[!htbp]
\begin{center}
\tikzset{every picture/.style={line width=0.75pt}} 

\begin{tikzpicture}[x=0.75pt,y=0.75pt,yscale=-1,xscale=1]

\draw  [color={rgb, 255:red, 74; green, 144; blue, 226 }  ,draw opacity=1 ][fill={rgb, 255:red, 208; green, 2; blue, 27 }  ,fill opacity=0.15 ][line width=1.5]  (103,53.6) -- (549,53.6) -- (549,255.6) -- (103,255.6) -- cycle ;
\draw [color={rgb, 255:red, 65; green, 117; blue, 5 }  ,draw opacity=1 ][line width=1.5]    (180,83.6) .. controls (162.27,126.94) and (191.11,142.14) .. (180.51,202.8) ;
\draw [shift={(180,205.6)}, rotate = 280.78] [color={rgb, 255:red, 65; green, 117; blue, 5 }  ,draw opacity=1 ][line width=1.5]    (14.21,-4.28) .. controls (9.04,-1.82) and (4.3,-0.39) .. (0,0) .. controls (4.3,0.39) and (9.04,1.82) .. (14.21,4.28)   ;
\draw [color={rgb, 255:red, 208; green, 2; blue, 27 }  ,draw opacity=1 ][line width=1.5]    (470,83.6) .. controls (452.27,126.94) and (481.11,142.14) .. (470.51,202.8) ;
\draw [shift={(470,205.6)}, rotate = 280.78] [color={rgb, 255:red, 208; green, 2; blue, 27 }  ,draw opacity=1 ][line width=1.5]    (14.21,-4.28) .. controls (9.04,-1.82) and (4.3,-0.39) .. (0,0) .. controls (4.3,0.39) and (9.04,1.82) .. (14.21,4.28)   ;
\draw [color={rgb, 255:red, 65; green, 117; blue, 5 }  ,draw opacity=1 ][line width=1.5]    (208,176.6) .. controls (224.58,151.25) and (196.47,131.6) .. (211.74,102.83) ;
\draw [shift={(213,100.6)}, rotate = 120.96] [color={rgb, 255:red, 65; green, 117; blue, 5 }  ,draw opacity=1 ][line width=1.5]    (14.21,-4.28) .. controls (9.04,-1.82) and (4.3,-0.39) .. (0,0) .. controls (4.3,0.39) and (9.04,1.82) .. (14.21,4.28)   ;
\draw [color={rgb, 255:red, 208; green, 2; blue, 27 }  ,draw opacity=1 ][line width=1.5]    (499,176.6) .. controls (515.58,151.25) and (487.47,131.6) .. (502.74,102.83) ;
\draw [shift={(504,100.6)}, rotate = 120.96] [color={rgb, 255:red, 208; green, 2; blue, 27 }  ,draw opacity=1 ][line width=1.5]    (14.21,-4.28) .. controls (9.04,-1.82) and (4.3,-0.39) .. (0,0) .. controls (4.3,0.39) and (9.04,1.82) .. (14.21,4.28)   ;

\draw (135,131.4) node [anchor=north west][inner sep=0.75pt]  [font=\Large]  {$h_{12}$};
\draw (398,132.4) node [anchor=north west][inner sep=0.75pt]  [font=\Large]  {$\textcolor[rgb]{0.82,0.01,0.11}{h_{ii} /\phi }$};
\draw (236,213.4) node [anchor=north west][inner sep=0.75pt]  [font=\LARGE]  {$horizon( r=r_{h})$};
\draw (112,209.4) node [anchor=north west][inner sep=0.75pt]  [font=\LARGE]  {$\eta \sim P_{absor}^{12}$};
\draw (426,203.4) node [anchor=north west][inner sep=0.75pt]  [font=\LARGE]  {$\textcolor[rgb]{0.82,0.01,0.11}{\zeta \sim P_{absor}^{ii/\phi }}$};

\end{tikzpicture}

\end{center}
    \caption{The figure shows schematically the relationship between the shear viscosity ($\eta$) and the absorption of the graviton h$_{12}$ (left side) and between the bulk viscosity ($\zeta$) and the absorption of a mixture of the graviton h$_{ii}$ and the scalar $\phi$.}
    \label{fig:visc}
\end{figure}

In this work we will present a talk about the techniques used to calculate the shear and bulk viscosities in gravity duals of Einstein \cite{Meyer:2007dy,Kharzeev:2007wb,Policastro:2001yc,Buchel:2007mf,Karsch:2007jc,Gubser:2008yx,Gubser:2008sz} and Horndeski \cite{Bravo-Gaete:2020lzs,Bravo-Gaete:2022lno,Santos:2023flb,Bravo-Gaete:2023iry,Santos:2023mee} at finite temperature involving a single scalar. Recent investigations \cite{Bravo-Gaete:2020lzs,Bravo-Gaete:2022lno,Santos:2023flb,Bravo-Gaete:2023iry,Santos:2023mee}, in modified gravity scenarios, the entropy viscosity relations $\eta/S$ and $\zeta/S$ can be extracted using Kubo's formula. In both the Anti-de Sitter/Conformal Field Theory (AdS/BCFT) correspondence and Anti-de Sitter/Boundary Conformal Field Theory (AdS/BCFT) correspondence scenarios \cite{Bravo-Gaete:2020lzs,Bravo-Gaete:2022lno,Santos:2023flb,Bravo-Gaete:2023iry,Santos:2023mee,Feng:2015oea}, we can use the strategy, illustrated in Fig \ref{fig:visc}; we start with a static, infinite thermal background, linearize the equations of motion around it, and solve these linearized equations in some appropriate approximation. This approximation is the long-wavelength, low-frequency one where we have the hydrodynamic effects.

\section{AdS/CFT versus AdS/BCFT}

We now know that the AdS/CFT correspondence postulates the equivalence between a certain gravitational theory and a non-gravitational theory of lower dimension \cite{Maldacena:1997re,Witten:1998qj}. Once established, this theory has led to new fields of research in recent years, ranging from holographic superconductors to applications in condensed matter systems \cite{Hartnoll:2008vx,Hartnoll:2008kx,Horowitz:2011dz}. Through the AdS/CFT correspondence, we know that IR divergences in the gravity side correspond to the UV divergences in CFT boundary theory. This relation is the IR-UV connection; see Fig. \ref{BTZ2}. 

\begin{figure}[!ht]

\begin{center}

\tikzset{every picture/.style={line width=0.75pt}} 

\begin{tikzpicture}[x=0.75pt,y=0.75pt,yscale=-1,xscale=1]

\draw  [fill={rgb, 255:red, 208; green, 2; blue, 27 }  ,fill opacity=0.43 ][line width=2.25]  (380,205) .. controls (380,199.48) and (384.48,195) .. (390,195) .. controls (395.52,195) and (400,199.48) .. (400,205) .. controls (400,210.52) and (395.52,215) .. (390,215) .. controls (384.48,215) and (380,210.52) .. (380,205) -- cycle ;
\draw  [fill={rgb, 255:red, 189; green, 16; blue, 224 }  ,fill opacity=0.79 ][line width=2.25]  (206,205) .. controls (206,199.48) and (210.48,195) .. (216,195) .. controls (221.52,195) and (226,199.48) .. (226,205) .. controls (226,210.52) and (221.52,215) .. (216,215) .. controls (210.48,215) and (206,210.52) .. (206,205) -- cycle ;
\draw  [fill={rgb, 255:red, 80; green, 227; blue, 194 }  ,fill opacity=0.51 ][line width=2.25]  (308,91) .. controls (308,85.48) and (312.48,81) .. (318,81) .. controls (323.52,81) and (328,85.48) .. (328,91) .. controls (328,96.52) and (323.52,101) .. (318,101) .. controls (312.48,101) and (308,96.52) .. (308,91) -- cycle ;
\draw [line width=1.5]    (226,205) -- (377,205) ;
\draw [shift={(380,205)}, rotate = 180] [color={rgb, 255:red, 0; green, 0; blue, 0 }  ][line width=1.5]    (14.21,-4.28) .. controls (9.04,-1.82) and (4.3,-0.39) .. (0,0) .. controls (4.3,0.39) and (9.04,1.82) .. (14.21,4.28)   ;
\draw [line width=1.5]    (310,98) -- (225,192.77) ;
\draw [shift={(223,195)}, rotate = 311.89] [color={rgb, 255:red, 0; green, 0; blue, 0 }  ][line width=1.5]    (14.21,-4.28) .. controls (9.04,-1.82) and (4.3,-0.39) .. (0,0) .. controls (4.3,0.39) and (9.04,1.82) .. (14.21,4.28)   ;
\draw [line width=1.5]    (318,101) .. controls (307.11,128.72) and (293.28,170.16) .. (375.48,196.22) ;
\draw [shift={(378,197)}, rotate = 197.01] [color={rgb, 255:red, 0; green, 0; blue, 0 }  ][line width=1.5]    (14.21,-4.28) .. controls (9.04,-1.82) and (4.3,-0.39) .. (0,0) .. controls (4.3,0.39) and (9.04,1.82) .. (14.21,4.28)   ;
\draw [line width=1.5]    (313,99) .. controls (302.06,126.86) and (254.48,181.45) .. (371.23,198.74) ;
\draw [shift={(373,199)}, rotate = 188.13] [color={rgb, 255:red, 0; green, 0; blue, 0 }  ][line width=1.5]    (14.21,-4.28) .. controls (9.04,-1.82) and (4.3,-0.39) .. (0,0) .. controls (4.3,0.39) and (9.04,1.82) .. (14.21,4.28)   ;

\draw (172,198) node [anchor=north west][inner sep=0.75pt]   [align=left] {CFT};
\draw (331,74) node [anchor=north west][inner sep=0.75pt]   [align=left] {UV-CFT};
\draw (403,194) node [anchor=north west][inner sep=0.75pt]   [align=left] {IR-CFT};

\end{tikzpicture}

\caption{Organized scheme of CFT space \cite{Santos:2024zoh}.}
\label{BTZ2}
\end{center}
\end{figure}

One of the most notable applications of the AdS/CFT correspondence has been describing strongly coupled systems nonperturbatively \cite{Ballon-Bayona:2022uyy}. Thus, this duality in a supersymmetric extension of the 4-dimensional Yang-Mills theory; although this is different from real QCD, it has provided some macroscopic properties at finite temperature that are close to those of real QCD \cite{Karsch:2007jc,Gubser:2008yx,Gubser:2008sz}. One of the most important applications found in recent years has been the ratio between shear viscosity and entropy density ($\eta/S$) \cite{Policastro:2001yc,Policastro:2002se}. This is very close to the result expected for the quark-gluon plasma observed in heavy ion collisions. Beyond the shear viscosity/entropy density ratio, another ratio is the bulk viscosity/entropy density ratio ($\zeta/S$) of this liquid is extremely low, which indicates that the plasma has the nature of a liquid and not a gas \cite{Ballon-Bayona:2022uyy}. A new way to probe these relations is by extending the AdS/BCFT correspondence with Horndeski gravity where it is possible to obtain small values of $\eta/S$, and $\zeta/S$ \cite{Santos:2023flb,Santos:2023mee}. This new duality \cite{Takayanagi:2011zk,Fujita:2011fp,dosSantos:2022scy,Santos:2024cwf} is an extension of the AdS/CFT correspondence \cite{Maldacena:1997re}, defining in the CFT a boundary in $d$-dimensional variety $ \mathcal{M}$ for an asymptotically $d+ 1$-dimensional AdS space $\mathcal{N}$ such that $\partial \mathcal{N}=\mathcal{M}~\cup~Q$, where $Q$ is a $d$-dimensional manifold satisfying $ \partial {Q}~\cap~\partial \mathcal{M}=\mathcal{P}$ (Fig. \ref{AdSBCFT}).

\begin{figure}[!htbp]
\begin{center}
   \tikzset{every picture/.style={line width=0.75pt}} 

\begin{tikzpicture}[x=0.75pt,y=0.75pt,yscale=-1,xscale=1]

\draw  [draw opacity=0][fill={rgb, 255:red, 208; green, 2; blue, 27 }  ,fill opacity=0.31 ][line width=1.5]  (384.5,109.01) .. controls (384.51,109.33) and (384.51,109.64) .. (384.52,109.95) .. controls (385.45,165.32) and (347.86,210.84) .. (300.57,211.63) .. controls (253.27,212.43) and (214.18,168.19) .. (213.26,112.82) .. controls (213.25,112.45) and (213.25,112.09) .. (213.24,111.72) -- (298.89,111.39) -- cycle ; \draw  [line width=1.5]  (384.5,109.01) .. controls (384.51,109.33) and (384.51,109.64) .. (384.52,109.95) .. controls (385.45,165.32) and (347.86,210.84) .. (300.57,211.63) .. controls (253.27,212.43) and (214.18,168.19) .. (213.26,112.82) .. controls (213.25,112.45) and (213.25,112.09) .. (213.24,111.72) ;  
\draw  [fill={rgb, 255:red, 80; green, 227; blue, 194 }  ,fill opacity=0.98 ][line width=1.5]  (253.05,81.7) .. controls (293.27,71.69) and (346.55,77.39) .. (372.05,94.42) .. controls (397.55,111.45) and (385.63,133.38) .. (345.41,143.39) .. controls (305.2,153.4) and (251.92,147.71) .. (226.42,130.68) .. controls (200.91,113.64) and (212.84,91.72) .. (253.05,81.7) -- cycle ;
\draw  [fill={rgb, 255:red, 80; green, 227; blue, 194 }  ,fill opacity=1 ] (246.2,158.2) .. controls (250.2,148.2) and (299.2,157.2) .. (287.2,162.2) .. controls (275.2,167.2) and (273.2,172.2) .. (270.2,177.2) .. controls (267.2,182.2) and (258.2,199.2) .. (249.2,176.2) .. controls (240.2,153.2) and (242.2,168.2) .. (246.2,158.2) -- cycle ;

\draw (380,158.4) node [anchor=north west][inner sep=0.75pt]  [font=\Large]  {$\partial Q$};
\draw (286,94.4) node [anchor=north west][inner sep=0.75pt]  [font=\LARGE]  {$\mathcal{M}$};
\draw (292,165.4) node [anchor=north west][inner sep=0.75pt]  [font=\Large]  {$Q$};
\draw (386,90.4) node [anchor=north west][inner sep=0.75pt]  [font=\LARGE]  {$\mathcal{P}$};
\draw (252,157) node [anchor=north west][inner sep=0.75pt]  [font=\large]  {$\mathcal{N}$};

\end{tikzpicture}
 \end{center}
    \caption{AdS/CFT correspondence in the presence of boundary hypersurface $Q$.}
    \label{AdSBCFT}   
\end{figure}
The corrections due to the extension of the AdS/CFT correspondence with Horndeski gravity provided exciting results regarding the Hawking-Page phase transition, the response of materials in paramagnetic/ferromagnetic transition to an external field, as well as magnetized plasma \cite{Santos:2023flb,Santos:2023mee,Santos:2021orr}. Thus, this lecture aims to revisit these results that agree with experiments and well-known results in the literature. Thus, Horndeski gravity in the AdS/BCFT scenario is a reasonable extension, since it conjectures scalar-tensor theories whose dynamics and solution space are relevant to the holographic duals at the boundary; Horndeski gravity has been shown in recent years to be reasonable for this purpose \cite{Caceres:2023gfa,Santos:2023eqp,Santos:2022lxj,DosSantos:2022exb,Santos:2020xox,Zhang:2022hxl}.
\section{Bulk viscosity and shear viscosity: from Einstein gravity to Horndeski gravity}
From a theoretical point of view, the zero expected value of the energy-momentum stress tensor ($\langle T^\alpha_{\ \ \alpha}\rangle$) reveals that the bulk viscosity is zero, describing a fluid according to \cite{Ballon-Bayona:2022uyy}. However, this only occurs in the Einstein gravity scenario; when we extend this scenario to Horndeski gravity, the expected value $\langle T^\alpha_{\ \ \alpha}\rangle$ is not zero \cite{Santos:2023flb,Santos:2023mee,Santos:2021orr}. The action of the Horndeski gravity is given by
\begin{eqnarray}\label{açao}
S_{bulk}&=&\int_{\mathcal{N}}d^{5} x \sqrt{-g}\; \left(\kappa {\cal L}_{\rm H}+\kappa {\cal L}_{\mathcal{M}}+{\cal L}_{mat}\right),\label{1}\\
{\cal L}_{\rm H}&=&(R-2\Lambda)\label{L1} -\frac{1}{2}(\alpha g_{\mu\nu}-\gamma\,  G_{\mu\nu})\nabla^{\mu}\phi\nabla^{\nu}\phi
\end{eqnarray}
Here, for ${\cal L}_{\rm H}$, we have that $R=g^{\mu \nu} R_{\mu \nu}$, $G_{\mu \nu}$ and $\Lambda$ represent the scalar curvature, the Einstein tensor, and the cosmological constant respectively, while that $\phi=\phi(r)$ is a scalar field, $\alpha$, and $\gamma$ are coupling constants where $\kappa={1}/{(16 \pi G_N)}$, with $G_N$ is the Newton Gravitational constant.
Under this scenario, to establish the $\mbox{AdS}_{5}/\mbox{BCFT}_{4}$ correspondence, we need to construct the terms of the boundary. Following the Refs. \cite{Santos:2023flb,Santos:2021orr}, these expressions are given by 
\begin{eqnarray}
S_{BCFT}&=&{2\kappa\int_{Q}{d^{4}x\sqrt{-h}\mathcal{L}_{bdry}}+2\int_{Q}{d^{4}x\sqrt{-h}\mathcal{L}_{mat}}+2\kappa\int_{ct}{d^{4}x\sqrt{-h}\mathcal{L}_{ct}}}\nonumber\\
&+&S^{Q}_{mat},\label{eq:BCFT}
\end{eqnarray}
with
\begin{eqnarray}
\mathcal{L}_{bdry}&=&(K-\Sigma)-\frac{\gamma}{4}(\nabla_{\mu}\phi\nabla_{\nu}\phi n^{\mu}n^{\nu}-(\nabla \phi)^2)K-\frac{\gamma}{4}\nabla_{\mu}\phi\nabla_{\nu}\phi K^{\mu\nu},\label{eq:Lbound}\\
\mathcal{L}_{ct}&=&c_{0}+c_{1}R+c_{2}R^{ij}R_{ij}+c_{3}R^{2}+b_{1}(\partial_{i}\phi\partial^{i}\phi)^{2}+\cdots.
\end{eqnarray}
For the Lagrangian $\mathcal{L}_{bdry}$, $K_{\mu\nu}=h^{\phantom{\mu}\beta}_{\mu}\nabla_{\beta}n_{\nu}$ corresponds to the extrinsic curvature where $K=h^{\mu\nu}K_{\mu\nu}$ is the trace, $h_{\mu\nu}$ is the induced metric while that $n^{\mu}$ is an outward pointing unit normal vector to the boundary of the hypersurface $Q$. Additionally, $\Sigma$ is the boundary tension on $Q$ and $S^{Q}_{mat}$ is the matter action on $Q$. ${\cal L}_{ct}$ represents the boundary counterterms, which do not influence the bulk dynamics and hence will be disregarded. The expected value of the energy-momentum stress tensor ($\langle T^\alpha_{\ \ \alpha}\rangle$) is given by
\begin{equation}
   \langle T^a_{\ \ a}\rangle = \epsilon - 3p = 4 \Omega + TS\,.
\end{equation}
where the entropy for a five-dimensional black hole
\begin{eqnarray}\label{ansatz}
&&ds^2= \frac{L^2}{r^2}\left(-f(r)\,dt^2+dx^2+dy^2+dw^2+\frac{dr^2}{f(r)}\right),\\
&&f(r)= 1-\left(\frac{r}{r_{h}}\right)^{4}
\end{eqnarray}
is given by \cite{Santos:2023mee}:
\begin{eqnarray}
S_{\rm bulk}&=&\frac{L^{2}V}{4r^{3}_{h}G_{N}}\left(1-\frac{\xi}{4}\right),\label{eq:Entbulk}\\
S_{\rm bdry}&=&\frac{L^{2}\Delta\,y_{Q}}{G_{N}}\left(1-\frac{\xi}{4}\right)\left(\frac{\xi\,L^{2}b(\theta{'})}{5r^{3}_{h}}-\frac{q(\theta^{'})}{3r_{h}}\right)\nonumber\\
&-&\frac{L^{2}\sec(\theta{'})\Delta\,y_{Q}}{G_{N}}\left(\frac{\xi\,L^{2}b(\theta{'})}{5r^{2}_{h}}-\frac{q(\theta^{'})}{2}\right).\label{BT8}
\end{eqnarray}
The meaning behind this overall entropy (\ref{BT8}) aligns with the Bekenstein-Hawking expression 
\begin{eqnarray}
S_{BH}=\frac{A}{4G_{N}}\label{BT9}\,,
\end{eqnarray}
where, in this case, the total area $A$ reads
\begin{eqnarray}
A&=&\frac{L^{2}V}{2r^{3}_{h}}\left(1-\frac{\xi}{4}\right)+4L^{2}\Delta\,y_{Q}\left(1-\frac{\xi}{4}\right)\left(\frac{\xi\,L^{2}b(\theta{'})}{5r^{3}_{h}}-\frac{q(\theta^{'})}{3r_{h}}\right)\nonumber\\
&-&4L^{2}\sec(\theta{'})\Delta\,y_{Q}\left(\frac{\xi\,L^{2}b(\theta{'})}{5r^{2}_{h}}-\frac{q(\theta^{'})}{2}\right),\,\label{BT10}
\end{eqnarray}
allowing us to obtain new contribution terms. For the sake of completeness, for the entropy $S$, we have that the information storage (delimited via the area $A$) increases as the magnitude of $\xi=\dfrac{\alpha+\gamma\Lambda}{\alpha}$ increases, as long as $\xi<0$.
\begin{figure}[!ht]
\begin{center}
\includegraphics[scale=0.55]{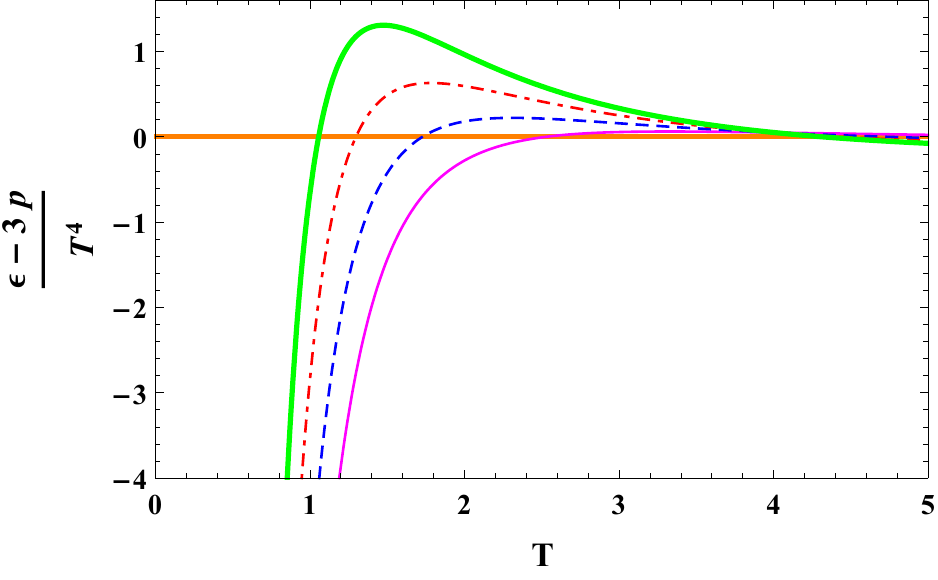}
\caption{Trace of the scaled energy-momentum tensor as a function of the temperature considering the following values $\kappa=1/4$, $\Lambda=-1$, $\alpha=8/3$ with $\gamma=0.1$ (solid line), $\gamma=0.2$ (dashed line), $\gamma=0.3$ (dot-dashed line), and $\gamma=0.4$ (thick line).}
\label{trace}
\end{center}
\end{figure}

For the above configurations, $\langle T^a_{\ \ a}\rangle$ is a non-zero expected value, suggesting the existence of residual information that does not cancel out at low temperatures as presented in \cite{Santos:2021orr}. The impact of this residual information in the form of entropy \cite{Santos:2023flb,Santos:2023mee,Santos:2021orr} for our results in QCD-like models are described by $\zeta/S$ and $\eta/S$

\begin{eqnarray}
&&\frac{\zeta}{S}=\frac{\sqrt{3}}{24\pi\mathcal{F}}\sqrt{\frac{\alpha+\gamma\Lambda}{\alpha-\gamma\Lambda}}\label{bulk},\\
&&\frac{\eta}{S}=\frac{1}{4\pi\mathcal{F}}\sqrt{\frac{3\alpha+\gamma\Lambda}{\alpha-\gamma\Lambda}},\label{visc}\\
&&\mathcal{F}=1+\frac{1}{T}\left(\frac{\xi\,L^2b(\theta{'})}{5}(4\pi T)^{3}-q(\theta^{'})\left(\frac{\pi T}{3}\right)\right)\nonumber\\
&-&\frac{\sec(\theta{'})}{\left(1-\frac{\xi}{4}\right)T}\left(-\frac{\xi\,L^2b(\theta{'})}{2}(\pi T)^{2}-\frac{q(\theta^{'})}{2}\right).
\end{eqnarray}

\begin{figure}[h!]
\begin{center}
\includegraphics[scale=0.6]{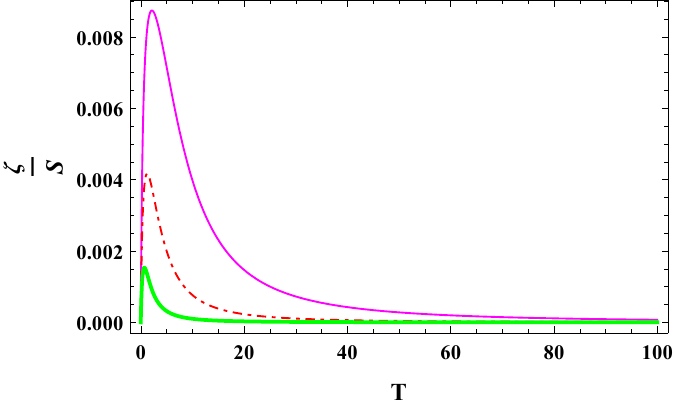} \includegraphics[scale=0.6]{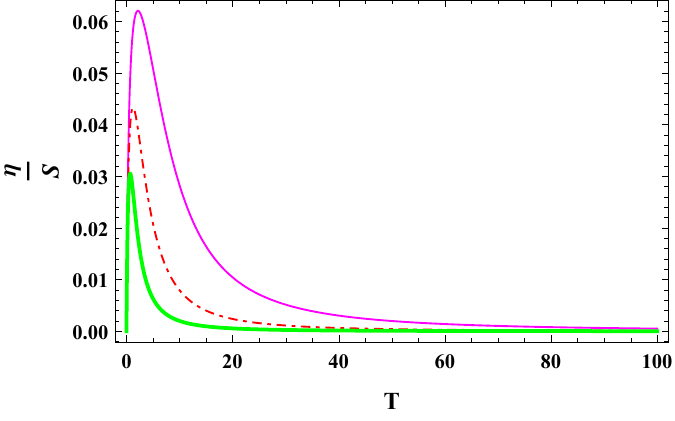}
\caption{{\bf Left panel:} the behavior of $\zeta/S$ with respect to $T$. {\bf Right panel:} The behavior of $\eta/S$ v/s $T$. For both situations, we consider $\alpha=8/3$, $\Lambda=-1$, $\gamma=1$ ({\sl pink curve}), $\gamma=2$ ({\sl red dot dashed curve}) and $\gamma=2.5$ ({\sl green thick curve}) \cite{Santos:2023mee}.}\label{zetaeta}
\end{center}
\end{figure}
Thus, as shown in Fig \ref{zetaeta} $\zeta/S$ and $\eta/S$ behave like $\langle T^\alpha_{\ \ \alpha}\rangle$ Fig \ref{trace}; they cancel each other out in the high-temperature region and are nonzero for low temperatures. In this way, the fluid achieves conformal behavior at high temperatures. The hydrodynamic pressures of the components of the holographic stress tensor in the Horndeski gravity scenario \cite{Santos:2023mee}. These components of the stress-energy tensor exhibit conformal hydrodynamic behavior near the boundary, that is, a region of high temperatures.

\section{Final remarks}
In this work, we explore aspects of the QGP quark-gluon plasma in a five-dimensional scenario using the AdS/BCFT within Horndeski gravity coupled to a single scalar, which contains the minimum amount of freedom necessary to match an equation of state with a family of black holes. This solution is the gravitational dual of a strongly coupled Yang-Mills plasma in BCFT$_{4}$, i.e., ($3+1$)-dimensions, with temperature $T$. We focus on studying the quantities $\zeta/S$, $\eta/S$ where $S$ is the entropy density in the quark-gluon plasma is essential for several reasons  \cite{Kharzeev:2007wb,Buchel:2007mf,Karsch:2007jc,Gubser:2008yx}. As we know, the quark-gluon plasma is a state of matter that existed in the early universe, which is recreated in heavy ion collision experiments \cite{Kharzeev:2007jp,Fukushima:2008xe}. The bulk viscosity/entropy density ratio is responsible for the equilibrium of a fluid subject to small expansion or compression; the shear viscosity/entropy density ratio $\eta/S$ is a quantity that is associated with energy dissipation due to the relative movement of the fluid layers \cite{Torres-Rincon:2012sda,Mishra:2020onx}. 

Through total entropy $S$, these transport coefficients are essential to understanding the dynamics of the quark-gluon plasma and can be used to extract information about its properties. Thus, studying such transport coefficients allows us to derive an agreement between our theoretical results, which have similar behaviors to the experimental results of the QCD. For example, at high temperatures, the quark-gluon plasma exhibits a conformal behavior; that is, the quantities $\zeta/S$, $\eta/S$ in this regime tend to zero. However, we will show that at low temperatures, the effects of anisotropy can be captured by the residual entropy \cite{Santos:2021orr,Santos:2023flb}, making the break of conformal symmetry evident. Furthermore, this anisotropic effect is also observed by the transverse and longitudinal components of the pressures exhibited by this plasma regime as they become evident within the Horndeski gravity scenario in the presence of a magnetic field \cite{Santos:2023mee}. In this case, we have agreement with the phenomenologically expected results \cite{Bali:2014kia}. 



\begin{thebibliography}{99}

\bibitem{Skokov:2009qp}
V.~Skokov, A.~Y.~Illarionov and V.~Toneev,
{\it Estimate of the magnetic field strength in heavy-ion collisions},
Int. J. Mod. Phys. A \textbf{24}, 5925-5932 (2009)
doi:10.1142/S0217751X09047570
[arXiv:0907.1396 [nucl-th]].

\bibitem{Fukushima:2016xgg}
K.~Fukushima,
{\it Evolution to the quark\textendash{}gluon plasma},
Rept. Prog. Phys. \textbf{80}, no.2, 022301 (2017)
doi:10.1088/1361-6633/80/2/022301
[arXiv:1603.02340 [nucl-th]].

\bibitem{Meyer:2007dy}
H.~B.~Meyer,
{\it A Calculation of the bulk viscosity in SU(3) gluodynamics},
Phys. Rev. Lett. \textbf{100}, 162001 (2008)
doi:10.1103/PhysRevLett.100.162001
[arXiv:0710.3717 [hep-lat]].

\bibitem{Kharzeev:2007wb}
D.~Kharzeev and K.~Tuchin,
{\it Bulk viscosity of QCD matter near the critical temperature},
JHEP \textbf{09}, 093 (2008)
doi:10.1088/1126-6708/2008/09/093
[arXiv:0705.4280 [hep-ph]].


\bibitem{Ahn:2022azl}
Y.~Ahn, M.~Baggioli, K.~B.~Huh, H.~S.~Jeong, K.~Y.~Kim and Y.~W.~Sun,
{\it Holography and magnetohydrodynamics with dynamical gauge fields},
[arXiv:2211.01760 [hep-th]].

\bibitem{Ballon-Bayona:2022uyy}
A.~Ballon-Bayona, J.~P.~Shock and D.~Zoakos,
{\it Magnetising the $ \mathcal{N} $ = 4 Super Yang-Mills plasma},
JHEP \textbf{06}, 154 (2022)
doi:10.1007/JHEP06(2022)154
[arXiv:2203.00050 [hep-th]].

\bibitem{Gubser:2008px}
S.~S.~Gubser,
{\it Breaking an Abelian gauge symmetry near a black hole horizon},
Phys. Rev. D \textbf{78}, 065034 (2008)
doi:10.1103/PhysRevD.78.065034
[arXiv:0801.2977 [hep-th]].


\bibitem{Gubser:1996de}
S.~S.~Gubser, I.~R.~Klebanov and A.~W.~Peet,
{\it Entropy and temperature of black 3-branes},
Phys. Rev. D \textbf{54} (1996), 3915-3919
doi:10.1103/PhysRevD.54.3915
[arXiv:hep-th/9602135 [hep-th]].


\bibitem{Burgess:1999vb}
C.~P.~Burgess, N.~R.~Constable and R.~C.~Myers,
{\it The Free energy of N=4 superYang-Mills and the AdS/CFT correspondence},
JHEP \textbf{08} (1999), 017
doi:10.1088/1126-6708/1999/08/017
[arXiv:hep-th/9907188 [hep-th]].


\bibitem{Policastro:2001yc}
G.~Policastro, D.~T.~Son and A.~O.~Starinets,
{\it The Shear viscosity of strongly coupled N=4 supersymmetric Yang-Mills plasma},
Phys. Rev. Lett. \textbf{87} (2001), 081601
doi:10.1103/PhysRevLett.87.081601
[arXiv:hep-th/0104066 [hep-th]].


\bibitem{Policastro:2002se}
G.~Policastro, D.~T.~Son and A.~O.~Starinets,
{\it From AdS/CFT correspondence to hydrodynamics},
JHEP \textbf{09} (2002), 043
doi:10.1088/1126-6708/2002/09/043
[arXiv:hep-th/0205052 [hep-th]].

\bibitem{Buchel:2007mf}
A.~Buchel,
{\it Bulk viscosity of gauge theory plasma at strong coupling},
Phys. Lett. B \textbf{663}, 286-289 (2008)
doi:10.1016/j.physletb.2008.03.069
[arXiv:0708.3459 [hep-th]].

\bibitem{Karsch:2007jc}
F.~Karsch, D.~Kharzeev and K.~Tuchin,
{\it Universal properties of bulk viscosity near the QCD phase transition},
Phys. Lett. B \textbf{663}, 217-221 (2008)
doi:10.1016/j.physletb.2008.01.080
[arXiv:0711.0914 [hep-ph]].

\bibitem{Gubser:2008yx}
S.~S.~Gubser, A.~Nellore, S.~S.~Pufu and F.~D.~Rocha,
{\it Thermodynamics and bulk viscosity of approximate black hole duals to finite temperature quantum chromodynamics},
Phys. Rev. Lett. \textbf{101}, 131601 (2008)
doi:10.1103/PhysRevLett.101.131601
[arXiv:0804.1950 [hep-th]].

\bibitem{Gubser:2008sz}
S.~S.~Gubser, S.~S.~Pufu and F.~D.~Rocha,
{\it Bulk viscosity of strongly coupled plasmas with holographic duals},
JHEP \textbf{08}, 085 (2008)
doi:10.1088/1126-6708/2008/08/085
[arXiv:0806.0407 [hep-th]].

\bibitem{Bravo-Gaete:2020lzs}
M.~Bravo-Gaete and F.~F.~Santos,
{\it Complexity of four-dimensional hairy anti-de-Sitter black holes with a rotating string and shear viscosity in generalized scalar\textendash{}tensor theories},
Eur. Phys. J. C \textbf{82}, no.2, 101 (2022)
doi:10.1140/epjc/s10052-022-10064-y
[arXiv:2010.10942 [hep-th]].

\bibitem{Bravo-Gaete:2022lno}
M.~Bravo-Gaete, F.~F.~Santos and H.~Boschi-Filho,
{\it Shear viscosity from black holes in generalized scalar-tensor theories in arbitrary dimensions},
Phys. Rev. D \textbf{106}, no.6, 066010 (2022)
doi:10.1103/PhysRevD.106.066010
[arXiv:2201.07961 [hep-th]].

\bibitem{Santos:2023flb}
F.~F.~Santos, M.~Bravo-Gaete, O.~Sokoliuk and A.~Baransky,
{\it AdS/BCFT Correspondence and Horndeski Gravity in the Presence of Gauge Fields: Holographic Paramagnetism/Ferromagnetism Phase Transition},
Fortsch. Phys. \textbf{71}, no.12, 2300008 (2023)
doi:10.1002/prop.202300008
[arXiv:2301.03121 [hep-th]].

\bibitem{Bravo-Gaete:2023iry}
M.~Bravo-Gaete, L.~Guajardo and F.~F.~Santos,
{\it Exploring the shear viscosity in four-dimensional planar black holes beyond general relativity},
Phys. Rev. D \textbf{107}, no.10, 104032 (2023)
doi:10.1103/PhysRevD.107.104032
[arXiv:2303.07493 [hep-th]].

\bibitem{Santos:2023mee}
F.~F.~Santos, M.~Bravo-Gaete, M.~M.~Ferreira and R.~Casana,
{\it Magnetized AdS/BCFT Correspondence in Horndeski Gravity},
doi:10.1002/prop.202400088
[arXiv:2310.17092 [hep-th]].

\bibitem{Feng:2015oea}
X.~H.~Feng, H.~S.~Liu, H.~L\"u and C.~N.~Pope,
{\it Black Hole Entropy and Viscosity Bound in Horndeski Gravity},
JHEP \textbf{11}, 176 (2015)
doi:10.1007/JHEP11(2015)176
[arXiv:1509.07142 [hep-th]].

  \bibitem{Maldacena:1997re} 
  J.~M.~Maldacena,
  {\it The Large N limit of superconformal field theories and supergravity},
  Int.\ J.\ Theor.\ Phys.\  {\bf 38}, 1113 (1999)
  [Adv.\ Theor.\ Math.\ Phys.\  {\bf 2}, 231 (1998)]
    [hep-th/9711200].
 
\bibitem{Witten:1998qj}
E.~Witten,
{\it Anti-de Sitter space and holography},
Adv. Theor. Math. Phys. \textbf{2} (1998), 253-291
doi:10.4310/ATMP.1998.v2.n2.a2
[arXiv:hep-th/9802150 [hep-th]].

	\bibitem{Hartnoll:2008vx}
	S.~A.~Hartnoll, C.~P.~Herzog and G.~T.~Horowitz,
	{\it Building a Holographic Superconductor},
	Phys. Rev. Lett. \textbf{101}, 031601 (2008)
	doi:10.1103/PhysRevLett.101.031601
	[arXiv:0803.3295 [hep-th]].
	
	\bibitem{Hartnoll:2008kx}
	S.~A.~Hartnoll, C.~P.~Herzog and G.~T.~Horowitz,
	{\it Holographic Superconductors},
	JHEP \textbf{12}, 015 (2008)
	doi:10.1088/1126-6708/2008/12/015
	[arXiv:0810.1563 [hep-th]].

\bibitem{Horowitz:2011dz}
G.~T.~Horowitz, J.~E.~Santos and B.~Way,
{\it A Holographic Josephson Junction},
Phys. Rev. Lett. \textbf{106} (2011), 221601
[arXiv:1101.3326 [hep-th]].

\bibitem{Santos:2024zoh}
F.~F.~Santos and H.~Boschi-Filho,
{\it Holographic complexity and residual entropy of a rotating charged BTZ black hole within Horndeski gravity},
[arXiv:2407.10004 [hep-th]].

\bibitem{Takayanagi:2011zk} 
T.~Takayanagi,
``Holographic Dual of BCFT,''
Phys.\ Rev.\ Lett.\  {\bf 107}, 101602 (2011),
[arXiv:1105.5165 [hep-th]].

\bibitem{Fujita:2011fp} 
M.~Fujita, T.~Takayanagi and E.~Tonni,
``Aspects of AdS/BCFT,''
JHEP {\bf 1111}, 043 (2011),
[arXiv:1108.5152 [hep-th]].

\bibitem{dosSantos:2022scy}
F.~F.~dos Santos,
{\it AdS/BCFT correspondence and BTZ black hole within electric field},
JHAP \textbf{4}, no.1, 81-92 (2022)
doi:10.22128/jhap.2022.504.1018
[arXiv:2206.09502 [hep-th]].

\bibitem{Santos:2024cwf}
F.~F.~Santos and H.~Boschi-Filho,
{\it Geometric Josephson junction},
[arXiv:2407.10008 [hep-th]].

\bibitem{Santos:2021orr}
F.~F.~Santos, E.~F.~Capossoli and H.~Boschi-Filho,
{\it AdS/BCFT correspondence and BTZ black hole thermodynamics within Horndeski gravity},
Phys. Rev. D \textbf{104}, no.6, 066014 (2021)
doi:10.1103/PhysRevD.104.066014
[arXiv:2105.03802 [hep-th]].
\bibitem{Caceres:2023gfa}
N.~Caceres, C.~Corral, F.~Diaz and R.~Olea,
{it Holographic renormalization of Horndeski gravity},
JHEP \textbf{05}, 125 (2024)
doi:10.1007/JHEP05(2024)125
[arXiv:2311.04054 [hep-th]].

\bibitem{Santos:2023eqp}
F.~F.~Santos, B.~Pourhassan and E.~N.~Saridakis,
{\it de Sitter Versus Anti-de Sitter in Horndeski-Like Gravity},
Fortsch. Phys. \textbf{72}, no.3, 2300228 (2024)
doi:10.1002/prop.202300228
[arXiv:2305.05794 [hep-th]].

\bibitem{Santos:2022lxj}
F.~F.~Santos, O.~Sokoliuk and A.~Baransky,
{\it Holographic Complexity of Braneworld in Horndeski Gravity},
Fortsch. Phys. \textbf{71}, no.2-3, 2200141 (2023)
doi:10.1002/prop.202200141
[arXiv:2210.11596 [hep-th]].

\bibitem{DosSantos:2022exb}
F.~F.~Dos Santos,
{\it Entanglement entropy in Horndeski gravity},
JHAP \textbf{3}, no.1, 1-14 (2022)
doi:10.22128/jhap.2022.488.1015
[arXiv:2201.02500 [hep-th]].

\bibitem{Santos:2020xox}
F.~F.~Santos,
{\it Rotating black hole with a probe string in Horndeski Gravity},
Eur. Phys. J. Plus \textbf{135}, no.10, 810 (2020)
doi:10.1140/epjp/s13360-020-00805-x
[arXiv:2005.10983 [hep-th]].

\bibitem{Zhang:2022hxl}
D.~Zhang, G.~Fu, X.~J.~Wang, Q.~Pan and J.~P.~Wu,
{\it Transport properties in the Horndeski holographic two-currents model},
Eur. Phys. J. C \textbf{83}, no.4, 316 (2023)
doi:10.1140/epjc/s10052-023-11444-8
[arXiv:2211.07074 [hep-th]].

\bibitem{Kharzeev:2007jp}
D.~E.~Kharzeev, L.~D.~McLerran and H.~J.~Warringa,
{\it The Effects of topological charge change in heavy ion collisions: 'Event by event P and CP violation'},
Nucl. Phys. A \textbf{803} (2008), 227-253
doi:10.1016/j.nuclphysa.2008.02.298
[arXiv:0711.0950 [hep-ph]].

\bibitem{Fukushima:2008xe}
K.~Fukushima, D.~E.~Kharzeev and H.~J.~Warringa,
{\it The Chiral Magnetic Effect},
Phys. Rev. D \textbf{78} (2008), 074033
doi:10.1103/PhysRevD.78.074033
[arXiv:0808.3382 [hep-ph]].

\bibitem{Torres-Rincon:2012sda}
J.~M.~Torres-Rincon,
{\it Hadronic transport coefficients from effective field theories},
doi:10.1007/978-3-319-00425-9
[arXiv:1205.0782 [hep-ph]].

\bibitem{Mishra:2020onx}
A.~K.~Mishra,
{\it Exploring the self interacting dark matter properties from low redshift observations},
Eur. Phys. J. C \textbf{82}, no.11, 1060 (2022)
doi:10.1140/epjc/s10052-022-10907-8
[arXiv:2002.11652 [astro-ph.CO]].


\bibitem{Bali:2014kia}
G.~S.~Bali, F.~Bruckmann, G.~Endr\"odi, S.~D.~Katz and A.~Sch\"afer,
{\it The QCD equation of state in background magnetic fields},
JHEP \textbf{08}, 177 (2014)
doi:10.1007/JHEP08(2014)177
[arXiv:1406.0269 [hep-lat]].


\end{thebibliography}
\end{document}